\documentclass[pmlr,twocolumn,10pt]{jmlr} 
\usepackage{booktabs}
\usepackage{microtype}
\usepackage{graphicx}
\usepackage{float}
\usepackage{placeins}
\usepackage{colortbl}
\usepackage{listings} 
\usepackage{changepage} 
\usepackage{xurl}
\usepackage{ragged2e} 
\usepackage{subcaption}
\usepackage{multirow} 
\usepackage{array}  
\usepackage{adjustbox}
\usepackage{longtable}
\usepackage{xcolor}
\usepackage{siunitx}
\usepackage{tcolorbox}
\renewcommand{\arraystretch}{1.1}
\theorembodyfont{\upshape}
\theoremheaderfont{\scshape}
\theorempostheader{:}
\theoremsep{\newline}

\usepackage{soul} 
\usepackage{xcolor} 

\usepackage{colortbl}
\newcommand{\rot}[1]{\rotatebox{60}{\parbox{2.5cm}{\centering\scriptsize #1}}}

\jmlryear{2026}
\jmlrvolume{297}
\jmlrsubmitted{} 
\jmlrpublished{} 
\jmlrworkshop{Conference on Health, Inference, and Learning (CHIL) 2026}

\title[Adaptive Test-Time Scaling for Zero-Shot Respiratory Audio Classification]{Adaptive Test-Time Scaling for Zero-Shot Respiratory Audio Classification}

\author{%
\Name{Tsai-Ning Wang} \Email{t.n.wang@tue.nl}\\
\addr Eindhoven University of Technology, The Netherlands
\AND
\Name{Herman Teun den Dekker} \Email{h.dendekker@erasmusmc.nl}\\
\addr Erasmus University Medical Center, The Netherlands
\AND
\Name{Lin-Lin Chen} \Email{L.Chen@tue.nl}\\
\addr Eindhoven University of Technology, The Netherlands
\AND
\Name{Neil Zeghidour} \Email{neil@kyutai.org}\\
\addr Kyutai, France
\AND
\Name{Aaqib Saeed} \Email{a.saeed@tue.nl}\\
\addr Eindhoven University of Technology, The Netherlands \\
Eindhoven Artificial Intelligence Systems Institute, The Netherlands
}

\begin{document}

\maketitle


\begin{abstract}
Automated respiratory audio analysis promises scalable, non-invasive disease screening, yet progress is limited by scarce labeled data and costly expert annotation. Zero-shot inference eliminates task-specific supervision, but existing methods apply uniform computation to every input regardless of difficulty. We introduce TRIAGE, a tiered zero-shot framework that adaptively scales test-time compute by routing each audio sample through progressively richer reasoning stages: fast label-cosine scoring in a joint audio-text embedding space (Tier-L), structured matching with clinician-style descriptors (Tier-M), and retrieval-augmented large language model reasoning (Tier-H). A confidence-based router finalizes easy predictions early while allocating additional computation to ambiguous inputs, enabling nearly half of all samples to exit at the cheapest tier. Across nine respiratory classification tasks without task-specific training, TRIAGE achieves a mean AUROC of 0.744, outperforming prior zero-shot methods and matching or exceeding supervised baselines on multiple tasks. Our analysis show that test-time scaling concentrates gains where they matter: uncertain cases see up to 19\% relative improvement while confident predictions remain unchanged at minimal cost. 
\end{abstract}

\begin{figure}[t]
    \centering
    \includegraphics[width=1.10\columnwidth]{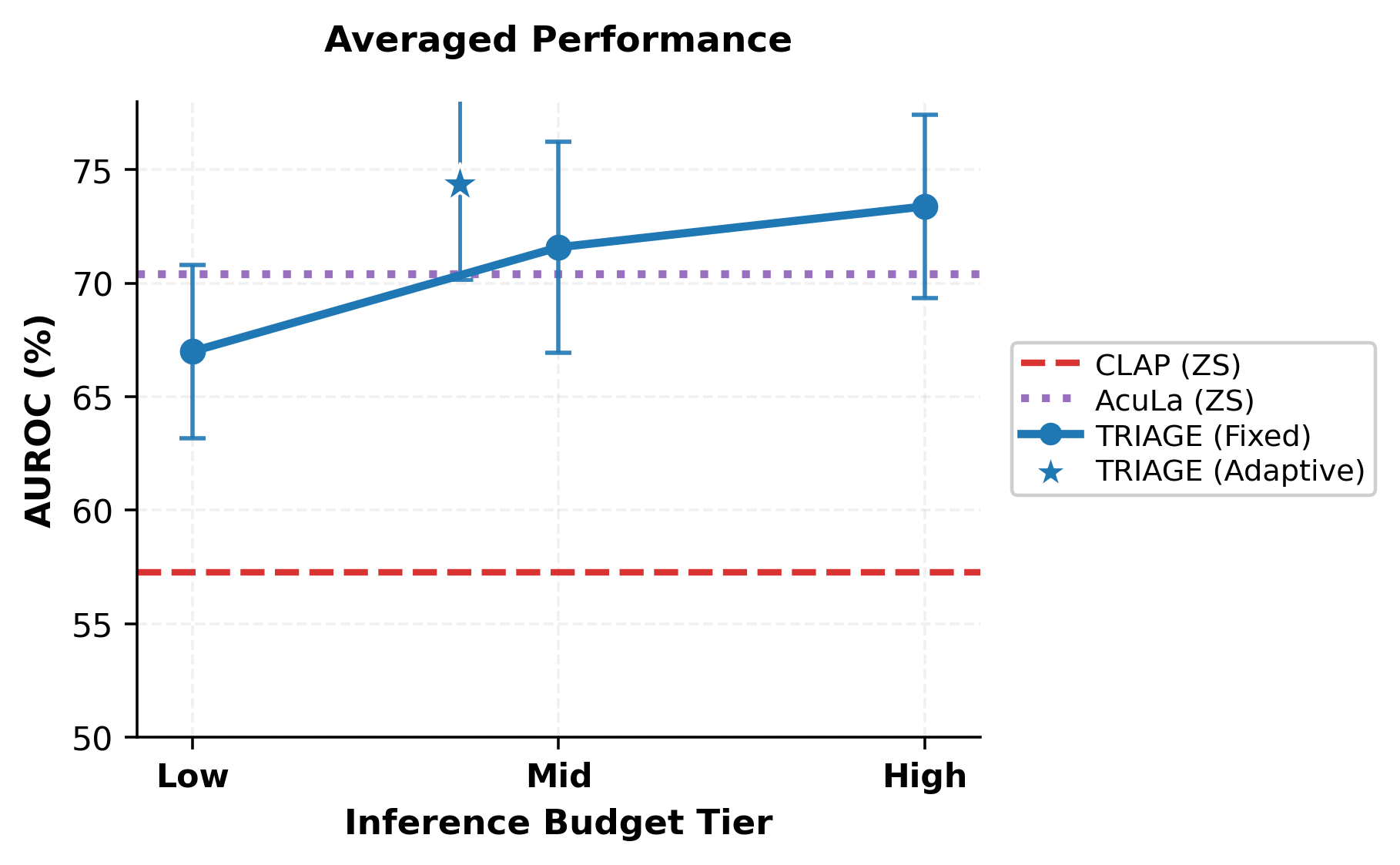}
    \captionsetup{font=small}
    \caption{Tiered test-time inference improves zero-shot respiratory audio classification. Points show mean AUROC (\%) averaged over 9 tasks versus budget tier. CLAP and AcuLa are zero-shot baselines (horizontal lines). TRIAGE (Fixed) evaluates three compute budgets by forcing all samples to run up to Tier-L/Tier-M/Tier-H (Low/Mid/High), while TRIAGE (Adaptive) reports the single operating point produced by confidence-based routing across tiers.}
    \label{fig:bar}
\end{figure}

{\small
\textbf{Data and Code Availability}\\
We use only publicly available datasets; the dataset descriptions and corresponding citations are provided in the Experiments section.  
Our source code is provided as anonymized supplemental material during review and will be made publicly available on GitHub upon acceptance.}

{\small
\textbf{Institutional Review Board (IRB)}\\
This research does not require IRB approval.
}

\section{Introduction}
\label{sec:intro}

\begin{figure*}[t]
    \centering
    \includegraphics[width=\textwidth]{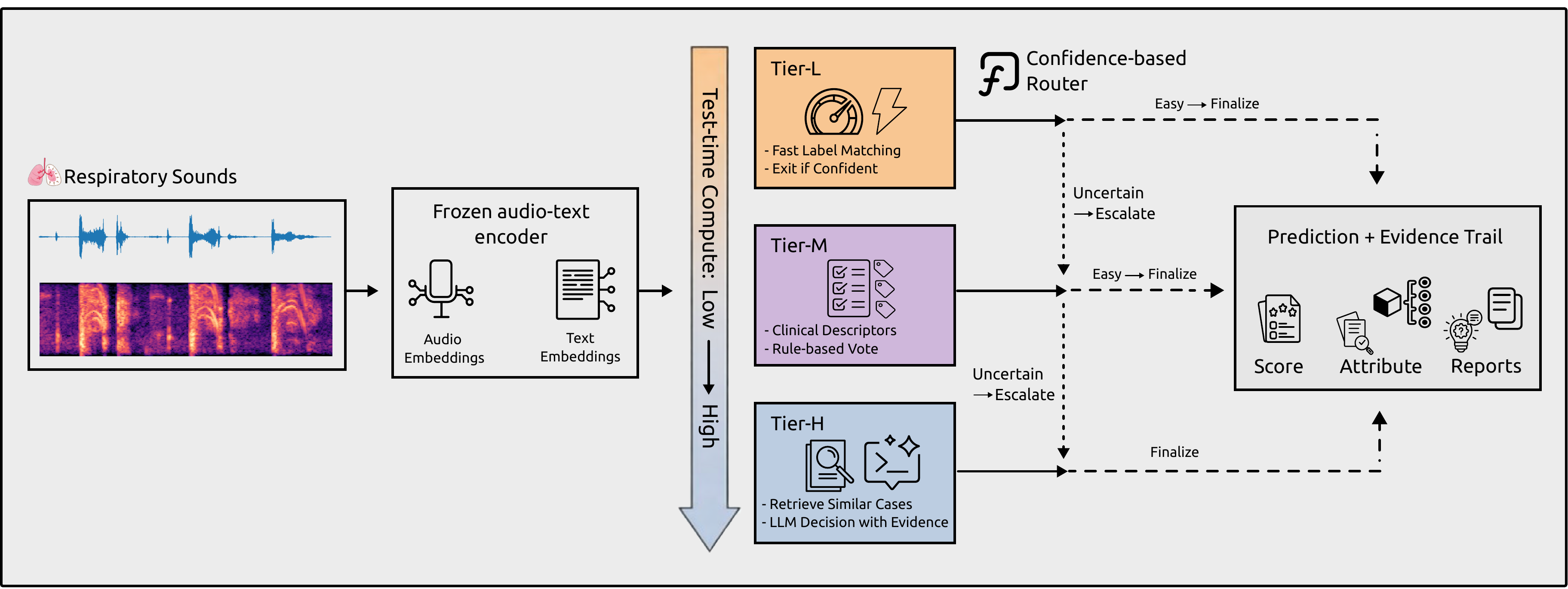}
\caption{\small{Overview of TRIAGE. A frozen audio--text encoder embeds recordings and medical text. \textbf{Tier-L} performs label cosine scoring with margin-based early exit; \textbf{Tier-M} matches clinical descriptors with rule voting; \textbf{Tier-H} retrieves nearest-neighbor reports and queries an off-the-shelf LLM, e.g., Gemini or GPT.}}

    \label{fig:tier_pipeline}
\end{figure*}

Automated analysis of heart and lung sounds increasingly relies on audio--text embedding models that project recordings into a shared latent space, where classification reduces to similarity scoring against label prompts or lightweight classifiers \citep{zhang2024towards, zhang2024respllm}. Yet current pipelines treat inference as a fixed-cost operation: a clean textbook wheeze and a faint crackle buried in motion artifact receive identical computation—one forward pass, one decision rule. This uniformity is mismatched to clinical reality. Auscultation recordings vary widely in signal quality, device characteristics, and diagnostic subtlety; some can be resolved from gross spectral features, while others require careful attention to timing within the respiratory cycle, the coexistence of multiple abnormalities, or acoustic cues that lie near the noise floor. When the same shallow inference is applied indiscriminately, difficult cases are under-served and overall robustness suffers.

Adapting models to each clinical environment through fine-tuning offers one remedy, but this path is often blocked: regulatory approval takes months, labeled data from the target site may be scarce or inaccessible, and computational resources are constrained. A more agile alternative is to improve inference itself—extracting more from a frozen model by spending additional computation selectively on recordings that need it.

This idea has gained traction in language and vision under the banner of \emph{test-time scaling}: sampling multiple candidate outputs, applying self-verification, or routing inputs through progressively richer reasoning stages based on difficulty \citep{snell2024scalingllmtesttimecompute, brown2024largelanguagemonkeysscaling, 2024archon}. These strategies complement the familiar benefits of scaling models and data at training time \citep{hestness2017deeplearningscalingpredictable, kaplan2020scalinglawsneurallanguage}, yet they have been studied almost exclusively for autoregressive generators on open-domain benchmarks. For embedding-based systems—particularly in safety-critical medical applications—test-time compute remains underexplored as a resource that can be allocated differentially across inputs without any task-specific supervision.

We address this gap with \textbf{TRIAGE} (\textbf{T}iered \textbf{R}etrieval and \textbf{I}nference for \textbf{A}udio with \textbf{G}ated \textbf{E}scalation), a three-tier inference framework for zero-shot respiratory audio classification. TRIAGE leaves the underlying audio--text encoder frozen and instead modulates computation per recording through gated escalation. In the first tier, each recording is scored against natural-language label prompts via cosine similarity; cases with a confident margin are finalized immediately. Recordings that fall below this confidence threshold advance to a second tier, where the audio is matched against clinician-approved descriptor templates—structured queries probing sound timing, quality, and anatomical location. A rule-based aggregator distills these matches into an interpretable attribute profile. The most ambiguous cases proceed to a third tier: nearest-neighbor retrieval over an external corpus of audio--report pairs supplies supporting evidence, and a large language model synthesizes a final prediction grounded in retrieved context. 

This architecture exposes an explicit compute--accuracy tradeoff. By tuning the confidence thresholds that govern tier transitions, practitioners can shift the operating point: resolve more cases cheaply when efficiency is paramount, or escalate aggressively when reliability matters most. Empirically, we evaluate TRIAGE on nine tasks spanning five public respiratory datasets. In a fully zero-shot setting—no fine-tuning, no labeled examples from the target data—TRIAGE improves over direct embedding similarity by a substantial margin on the majority of tasks, with the largest gains concentrated on recordings that require escalation. These results suggest that adaptive inference-time computation is a practical way for improving robustness in medical audio, independent of model or data scaling.

\vspace{0.5em}
\noindent We make the following contributions:
\begin{itemize}
    \item \textbf{Adaptive test-time scaling for medical audio embeddings.} We show that selectively increasing inference computation on uncertain recordings—without task-specific training—can substantially improve zero-shot classification, offering an alternative pathway to robustness when model adaptation is infeasible.
    \item \textbf{A compute-aware formulation of auscultation inference.} We cast respiratory audio classification as a staged decision process in which test-time computation is an explicit, tunable resource, enabling principled study of efficiency--accuracy tradeoffs across heterogeneous recording conditions.
    \item \textbf{An interpretable, clinically grounded inference pipeline.} TRIAGE decomposes classification into label scoring, descriptor-based attribute extraction, and retrieval-augmented reasoning. Each stage yields human-readable outputs—confidence scores, attribute profiles, retrieved evidence—providing transparency into why a recording was escalated and how the final decision was reached.
\end{itemize}

\section{Related Work}
\label{related_works}
Our work draws on three lines of research: domain-specific audio encoders for auscultation, audio--language models that enable zero-shot classification, and recent methods that treat inference-time computation as a tunable resource. We review each in turn, highlighting how TRIAGE builds on and departs from prior approaches.
\vspace{-0.8em}
\subsection{Domain-Specific Audio Encoders for Auscultation}
\vspace{-0.3em}
Early AI-aided auscultation used hand-crafted features or shallow CNNs trained on small, single-task datasets (e.g., murmur or wheeze detection, normal--abnormal screening). Such models were tightly coupled to their training distribution, and adapting to new devices, populations, or label spaces typically required retraining.
Recent foundation-style encoders pretrained on large, heterogeneous respiratory audio improve transfer: OPERA aggregates multi-source cough and breath corpora and shows strong transfer to held-out tasks \citep{zhang2024towards}, while AcuLa aligns respiratory audio with a medical language model to inject clinical semantics and achieve state-of-the-art results across broad benchmarks \citep{wang2025languagemodelssemanticteachers}. In contrast, TRIAGE treats pretrained representations as fixed and improves performance through test-time inference, targeting recordings where a single forward pass and fixed decision rule (e.g., a linear head) are insufficient.

\subsection{Audio--Language Models and Zero-Shot Classification}
A complementary strategy learns joint audio--text embeddings for zero-shot transfer via natural-language prompts. CLIP popularized this paradigm in vision by training dual encoders on image--caption pairs and classifying by matching inputs to textual label descriptions \citep{radford2021learningtransferablevisualmodels}. AudioCLIP and CLAP extend this idea to audio with contrastive audio--text alignment, enabling competitive zero-shot tagging and retrieval across diverse sound categories \citep{guzhov2021audioclipextendingclipimage, elizalde2022claplearningaudioconcepts}.

Beyond embedding models, larger audio--language systems combine acoustic front-ends with autoregressive decoders or instruction-following LLMs. In the clinical domain, RespLLM integrates respiratory audio with clinical text in a multimodal language model to improve robustness under dataset shift \citep{zhang2024respllm}. CaReAQA couples a self-supervised auscultation encoder with an LLM for open-ended diagnostic QA over medical sounds and introduces a cardiorespiratory audio QA benchmark that also evaluates transfer to closed-ended classification \citep{wang2025careaqacardiacrespiratoryaudio}. These works show that shared audio--language representations can support flexible and interpretable zero-shot behavior.
However, inference is typically uniform across recordings: inputs receive the same procedure and compute budget, whether via embedding similarity or generation. TRIAGE retains zero-shot audio--text matching but adds a gated pipeline that allocates more computation to ambiguous cases by escalating them to progressively richer reasoning stages.
\vspace{-0.8em}
\subsection{Test-Time Scaling and Adaptive Inference}
\vspace{-0.3em}
Recent work treats inference-time computation as a tunable budget rather than a fixed cost, complementing classical scaling laws that relate performance to model size and training compute \citep{kaplan2020scalinglawsneurallanguage, hestness2017deeplearningscalingpredictable}. With the model held fixed, additional test-time effort can still improve outcomes.
In language modeling, repeated sampling yields predictable accuracy gains as the number of candidate generations increases \citep{brown2024largelanguagemonkeysscaling}. Archon composes inference-time strategies (e.g., verification, critique, multi-agent debate) into pipelines that can outperform single-pass baselines under matched token budgets \citep{2024archon}, and recent theory begins to characterize when test-time scaling should succeed \citep{chen2025provablescalinglawstesttime}. For classification, TestNUC applies test-time consistency with nearest neighbors in embedding space \citep{zou-etal-2025-testnuc}. For dual encoders, decomposing inputs into fine-grained attribute comparisons can improve out-of-distribution retrieval without changing encoder weights \citep{xiao2024scalingttc}, and prompt engineering alone can unlock diverse behaviors from frozen encoders \citep{koukounas2024jinaclipclipmodel}.
TRIAGE brings these ideas to medical audio. We keep a pretrained auscultation encoder fixed and introduce a three-tier pipeline—label-name scoring, structured clinical-attribute matching, and retrieval-augmented LLM reasoning—that allocates more computation to ambiguous recordings. Our setting is embedding-based, safety-critical, and fully zero-shot: no gradient updates and no labeled examples from the target distribution.

\vspace{-0.8em}
\section{Methodology}
\label{sec:method}
\vspace{-0.3em}
We present TRIAGE, a three-tier inference framework for zero-shot auscultation classification. We allocate test-time computation adaptively: easy recordings stop early, ambiguous ones escalate to richer reasoning. All tiers share a frozen audio--text encoder; no task-specific parameters are learned.

\subsection{Problem Formulation}
\label{sec:problem}
Let $\mathcal{X}$ denote auscultation recordings and $\mathcal{Y}_j$ the label set for task $j$. 
Given a test recording $x\in\mathcal{X}$, we predict $\hat{y}\in\mathcal{Y}_j$ in a fully zero-shot setting (no parameter updates and no labeled examples from the target task).

We assume access to three frozen resources:
\begin{enumerate}
\item a frozen audio--text embedding model with encoders $f_{\text{audio}}$ and $f_{\text{text}}$;
\item a clinician-defined descriptor system (groups + text templates) and task-specific rule tables;
\item a retrieval corpus $\mathcal{R}$ of audio--report pairs used only for Tier-H retrieval.
\end{enumerate}

\subsection{Audio--Text Embedding Model}
\label{sec:embedding}

The backbone of TRIAGE is a frozen dual-encoder that embeds audio and text into a shared $d$-dimensional space. The audio encoder
\[
f_{\text{audio}}: \mathcal{X} \to \mathbb{R}^d, \quad x \mapsto \mathbf{a}=f_{\text{audio}}(x),
\]
maps a recording to a unit-normalized embedding, and the text encoder
\[
f_{\text{text}}: \mathcal{T} \to \mathbb{R}^d, \quad t \mapsto \mathbf{t}=f_{\text{text}}(t),
\]
maps a text string to a unit-normalized embedding. We score alignment by cosine similarity,
\[
s(\mathbf{a},\mathbf{t})=\mathbf{a}^\top \mathbf{t},
\]
and TRIAGE varies only the text queries and how similarities are aggregated.

\subsection{Clinical Attribute System}
\label{sec:attributes}

Direct matching between audio embeddings and class-name embeddings can be brittle when class names are short or ambiguous. We therefore define a structured attribute system with \(K\) descriptor groups
\[
\mathcal{G}=\{g_1,\ldots,g_K\}.
\]
Each group \(g_k\) targets a clinically meaningful dimension (e.g., timing, anatomic location, sound quality, pitch, adventitious findings) and contains mutually exclusive options
\[
\mathcal{O}_k=\{o_{k,1},\ldots,o_{k,M_k}\}, \quad |\mathcal{O}_k|=M_k.
\]
Each option \(o_{k,m}\) is associated with a natural-language template \(t_{k,m}\in\mathcal{T}\). The full descriptor set and templates (clinician-reviewed) are provided in Appendix~\ref{apd:descriptors}.

Given \(x\) with embedding \(\mathbf{a}=f_{\text{audio}}(x)\), we score every option by cosine similarity:
\[
s_{k,m}(x)=s\bigl(\mathbf{a},\, f_{\text{text}}(t_{k,m})\bigr), \quad k\in[K],\; m\in[M_k].
\]
We select the top option per group,
\[
m_k^{*}(x)=\arg\max_{m\in[M_k]} s_{k,m}(x),
\]
yielding a descriptor profile
\[
\mathbf{z}(x)=\bigl(o_{1,m_1^{*}},\ldots,o_{K,m_K^{*}}\bigr)\in \mathcal{O}_1\times\cdots\times\mathcal{O}_K.
\]

\paragraph{Rule-based label mapping.}
For each task \(j\) with label set \(\mathcal{Y}_j\), a fixed rule table
\[
\Phi_j:\mathcal{O}_1\times\cdots\times\mathcal{O}_K \to \mathbb{R}^{C_j}
\]
maps \(\mathbf{z}(x)\) to label scores, encoding which attribute configurations support each class. The rule tables require no learned parameters; an example is provided in Appendix~\ref{apd:descriptors}.

\subsection{Three-Tier Inference Pipeline}
\label{sec:three_tier}
TRIAGE processes each recording \(x\) through up to three stages (Figure~\ref{fig:tier_pipeline}): \textbf{Tier-L} (label cosine scoring), \textbf{Tier-M} (descriptor templates + rule voting), and \textbf{Tier-H} (kNN report retrieval + LLM decision).

\subsubsection{Tier-L: Label-Similarity Scoring}
\label{sec:tier1}
For each class \(y\in\mathcal{Y}_j\), we encode its name to obtain \(\mathbf{t}_y=f_{\text{text}}(\mathrm{name}(y))\). Given \(\mathbf{a}=f_{\text{audio}}(x)\), we compute
\[
s_y(x)=\mathbf{a}^{\top}\mathbf{t}_y,\quad y\in\mathcal{Y}_j,
\]
and predict
\[
\hat{y}^{(L)}(x)=\arg\max_{y\in\mathcal{Y}_j} s_y(x).
\]
We use the top-two margin as confidence:
\[
c_L(x)=s_{(1)}(x)-s_{(2)}(x).
\]

\subsubsection{Tier-M: Descriptor-Based Decision}
\label{sec:tier2}
Tier-M selects the highest-scoring template per descriptor group to form \(\mathbf{z}(x)=\{m_k^{*}\}_{k=1}^{K}\). A task-specific rule table converts \(\mathbf{z}(x)\) to label scores \(\{r_y(x)\}_{y\in\mathcal{Y}_j}\), predicting
\[
\hat{y}^{(M)}(x)=\arg\max_{y\in\mathcal{Y}_j} r_y(x),
\]
with routing confidence
\[
c_M(x)=r_{(1)}(x)-r_{(2)}(x).
\]

\subsubsection{Tier-H: Retrieval-Augmented LLM Reasoning}
\label{sec:tier3}
For remaining uncertain cases, we retrieve \(k\) nearest neighbors from a corpus \(\mathcal{R}\) of audio embeddings paired with clinician-authored reports:
\[
\mathcal{N}_k(x)=\operatorname{top\text{-}k}_{i}\; s\bigl(\mathbf{a},\mathbf{a}_i^{\mathcal{R}}\bigr).
\]
We prompt the LLM with \(\mathbf{z}(x)\), Tier-L scores \(\{s_y(x)\}\), and retrieved report snippets, and parse the final prediction:
\[
\hat{y}^{(H)}(x)=\textsc{Parse}\bigl(\textsc{LLM}(P(x))\bigr).
\]
The prompt template is in Appendix~\ref{apd:prompt-tierh}.

\subsection{Gated Escalation Policy}
\label{sec:escalation}

The three tiers are composed via a gated escalation policy parameterized by thresholds \(\tau_L\) and \(\tau_M\). Let \(c_L(x)\) and \(c_M(x)\) denote the confidence scores from Tier-L and Tier-M, respectively. The final prediction is:
\[
\hat{y}(x) =
\begin{cases}
\hat{y}^{(L)}(x) & \text{if } c_L(x) \geq \tau_L, \\[4pt]
\hat{y}^{(M)}(x) & \text{if } c_L(x) < \tau_L \text{ and } c_M(x) \geq \tau_M, \\[4pt]
\hat{y}^{(H)}(x) & \text{otherwise}.
\end{cases}
\]
This policy ensures that computation scales with difficulty: high-confidence recordings terminate early, while ambiguous cases receive the full inference budget.

\paragraph{Threshold selection.}
$\tau_L$ and $\tau_M$ trade accuracy for compute: lower thresholds escalate more cases, while higher thresholds finalize earlier. We select $(\tau_L,\tau_M)$ on the validation split and keep them fixed for test evaluation (Section~\ref{sec:test_time_setup}).

\paragraph{Computational cost.}
Let $T_L,T_M,T_H$ be per-recording costs and $\alpha_M,\alpha_H$ the fractions reaching Tier-M and Tier-H. The expected cost is
\[
\bar{T}=T_L+\alpha_M T_M+\alpha_H T_H.
\]
Since $T_H\gg T_M>T_L$, controlling $\alpha_H$ is the main lever; the gate reserves Tier-H for genuinely uncertain cases.

\section{Experiments}
\label{sec:experiments}

We evaluate TRIAGE on nine auscultation classification tasks spanning five public datasets. Our experiments assess zero-shot performance against supervised baselines, quantify the contribution of each tier, and ablate key design choices.

\subsection{Tasks and Datasets}
\label{sec:datasets}

We evaluate nine auscultation classification tasks drawn from five public corpora of respiratory sounds (details in Appendix~\ref{apd:datasets}). The tasks include COVID-19 detection from exhalation and cough (UK COVID-19; \citealp{Coppock2023_UKCOVID19VocalAudio_Zenodo}), COVID-19 and gender classification from crowdsourced coughs (CoughVID; \citealp{Orlandic2021_COUGHVID_Zenodo}), COPD-versus-healthy screening on stethoscope recordings (ICBHI; \citealp{DVN/HT6PKI_2023}), smoking status and gender classification from cough audio (Coswara; \citealp{Bhattacharya2023_Coswara}), obstructive-versus-healthy lung sound classification (KAUH; \citealp{Fraiwan2021_lungsounds_mendeley}), and five-class COPD severity grading (Resp.@TR; \citealp{AltanKutlu2020_respiratorydatabaseTR}). For corpora that provide official subject-level splits, we adopt them directly; otherwise, we construct subject-disjoint train/validation/test partitions with approximate ratios 60/20/20. The proposed zero-shot pipeline is evaluated on the test splits only.

\subsection{Embedding Model and Baselines}
\label{sec:backbone}

Our method assumes a frozen audio--text encoder that maps both auscultation audio and textual descriptions into a shared embedding space, as described in Section~\ref{sec:embedding}. In all experiments, we instantiate this backbone with AcuLa~\citep{wang2025languagemodelssemanticteachers}, a recent medical audio--language encoder, to provide a strong and reproducible foundation and enable direct comparison to prior zero-shot methods. Concretely, we instantiate \(f_{\text{audio}}\) as a domain-specific encoder pretrained on a large multi-dataset collection of heart and lung sounds with self-supervised objectives, and \(f_{\text{text}}\) as a medical language encoder trained on clinical text. AcuLa leverages an alignment stage in which structured metadata from existing auscultation datasets are converted into synthetic clinical reports and used to align audio embeddings with text embeddings via a symmetric contrastive loss, while keeping the language encoder frozen. After this alignment stage, both encoders are frozen and used unchanged for all experiments.

\subsection{Test-Time Inference Setup}
\label{sec:test_time_setup}

We follow the pipeline described in Section~\ref{sec:three_tier} and specify here the test-time thresholds and hyperparameters used in evaluation.

\paragraph{Tier-L configuration.}
For binary tasks with label set \(\mathcal{Y}_j = \{y^{(1)}, y^{(2)}\}\), we define the Tier-L confidence as the absolute margin:
\[
c_L(x) = \left| s(\mathbf{a}, \mathbf{t}_{y^{(1)}}) - s(\mathbf{a}, \mathbf{t}_{y^{(2)}}) \right|.
\]
For multi-class tasks, \(c_L(x)\) is the margin between the top two similarities, as defined in Section~\ref{sec:tier1}. We finalize at Tier-L when \(c_L(x) \geq \tau_L\) with \(\tau_L = 0.20\); otherwise, the recording is routed to Tier-M.

\paragraph{Tier-M configuration.}
Tier-M produces rule-based label scores via the task-specific rule table \(\Phi_j\). We define \(c_M(x)\) as the absolute difference between the top two label scores. The threshold \(\tau_M\) is selected on the validation split: we sweep a small set of candidate values (\(\tau_M \in \{0.04, 0.08, 0.12, 0.16, 0.20\}\)), finalize recordings with \(c_M(x) \geq \tau_M\), and choose the value that yields the best validation performance on the finalized subset. We fix \(\tau_M\) per task and apply it unchanged on the test split.

\paragraph{Tier-H configuration.}
Tier-H is invoked only when \(c_L(x) < \tau_L\) and \(c_M(x) < \tau_M\). We retrieve the top \(k = 3\) nearest neighbors from the retrieval corpus \(\mathcal{R}\) using FAISS~\citep{douze2025faisslibrary}, concatenate the retrieved clinical reports into a context block, and query Gemini 3 Pro ~\citep{gemini3pro_modelcard} with greedy decoding (\(T = 0\)) to obtain the final prediction. Appendix~\ref{apd:llm_ablation} reports an ablation over alternative LLM backends under identical retrieval and prompt settings.

\begin{table*}[t]
\centering
\caption{Performance comparison across respiratory audio classification tasks. AUROC scores ($\uparrow$) for supervised baselines (linear probing with frozen encoders), zero-shot methods, and our proposed zero-shot tiered inference policies. \textbf{Supervised baselines} require task-specific training data. \textbf{Zero-shot methods} operate without any task-specific training: CLAP uses text-audio similarity, while our tiered policies use Tier-L (label-score cosines), Tier-M (descriptor retrieval with rule voting), Tier-H (FAISS kNN report retrieval + LLM reasoning), and Adaptive (confidence-based hierarchical routing across all three tiers).}
\label{tab:policy_auroc}
\setlength{\tabcolsep}{2.5pt}
\renewcommand{\arraystretch}{1.30}
\small
\begin{adjustbox}{max width=\textwidth}
\begin{tabular}{@{}l *{9}{c}@{}}
\toprule
\textbf{Method} & \rot{UKCOV-EX-1} & \rot{UKCOV-CO-1} & \rot{CVID-CO-1} & \rot{CVID-CO-2} & \rot{ICBHI-LS-1} & \rot{COSW-CO-1} & \rot{COSW-CO-2} & \rot{KAUH-LS-1} & \rot{RESPTR-LS-1} \\
\midrule
\multicolumn{10}{l}{\textit{\textbf{Linear Probing Baselines} — (requires task-specific training)}} \\
\addlinespace[0.05cm]
VGGish & 0.580{\scriptsize$\pm$0.001} & 0.557{\scriptsize$\pm$0.005} & 0.538{\scriptsize$\pm$0.028} & 0.600{\scriptsize$\pm$0.001} & 0.605{\scriptsize$\pm$0.077} & 0.507{\scriptsize$\pm$0.027} & 0.606{\scriptsize$\pm$0.003} & 0.605{\scriptsize$\pm$0.036} & 0.590{\scriptsize$\pm$0.034} \\
AudioMAE & 0.549{\scriptsize$\pm$0.001} & 0.616{\scriptsize$\pm$0.001} & 0.554{\scriptsize$\pm$0.004} & 0.628{\scriptsize$\pm$0.001} & 0.886{\scriptsize$\pm$0.017} & 0.549{\scriptsize$\pm$0.022} & 0.724{\scriptsize$\pm$0.001} & 0.616{\scriptsize$\pm$0.041} & 0.510{\scriptsize$\pm$0.021} \\
CLAP & 0.565{\scriptsize$\pm$0.001} & 0.648{\scriptsize$\pm$0.003} & 0.599{\scriptsize$\pm$0.007} & 0.665{\scriptsize$\pm$0.001} & 0.933{\scriptsize$\pm$0.005} & 0.680{\scriptsize$\pm$0.009} & 0.742{\scriptsize$\pm$0.001} & 0.697{\scriptsize$\pm$0.004} & 0.636{\scriptsize$\pm$0.045} \\
OGT & 0.605{\scriptsize$\pm$0.001} & 0.677{\scriptsize$\pm$0.001} & 0.552{\scriptsize$\pm$0.003} & 0.735{\scriptsize$\pm$0.000} & 0.741{\scriptsize$\pm$0.011} & 0.650{\scriptsize$\pm$0.005} & 0.825{\scriptsize$\pm$0.001} & 0.703{\scriptsize$\pm$0.016} & 0.606{\scriptsize$\pm$0.015} \\
AcuLa & 0.698{\scriptsize$\pm$0.001} & 0.730{\scriptsize$\pm$0.008} & 0.887{\scriptsize$\pm$0.003} & 0.796{\scriptsize$\pm$0.004} & 0.826{\scriptsize$\pm$0.014} & 0.830{\scriptsize$\pm$0.011} & 0.845{\scriptsize$\pm$0.004} & 0.752{\scriptsize$\pm$0.019} & 0.710{\scriptsize$\pm$0.028} \\
\midrule
\midrule
\multicolumn{10}{l}{\textit{\textbf{Zero-Shot Methods}}} \\
\addlinespace[0.05cm]
CLAP & 0.528{\scriptsize$\pm$0.000} & 0.542{\scriptsize$\pm$0.000} & 0.540{\scriptsize$\pm$0.000} & 0.574{\scriptsize$\pm$0.000} & 0.687{\scriptsize$\pm$0.000} & 0.556{\scriptsize$\pm$0.000} & 0.608{\scriptsize$\pm$0.000} & 0.566{\scriptsize$\pm$0.000} & 0.552{\scriptsize$\pm$0.000} \\
AcuLa & 0.602{\scriptsize$\pm$0.000} & 0.665{\scriptsize$\pm$0.000} & 0.768{\scriptsize$\pm$0.000} & 0.683{\scriptsize$\pm$0.000} & 0.789{\scriptsize$\pm$0.000} & 0.755{\scriptsize$\pm$0.000} & 0.714{\scriptsize$\pm$0.000} & 0.702{\scriptsize$\pm$0.000} & 0.656{\scriptsize$\pm$0.000} \\
\midrule
\multicolumn{10}{l}{\textit{\textbf{TRIAGE (Ours)}}} \\
\addlinespace[0.05cm]
Tier-L & 0.593{\scriptsize$\pm$0.000} & 0.627{\scriptsize$\pm$0.000} & 0.722{\scriptsize$\pm$0.000} & 0.668{\scriptsize$\pm$0.000} & 0.706{\scriptsize$\pm$0.000} & 0.717{\scriptsize$\pm$0.000} & 0.716{\scriptsize$\pm$0.000} & 0.670{\scriptsize$\pm$0.000} & 0.610{\scriptsize$\pm$0.000} \\
Tier-M & 0.690{\scriptsize$\pm$0.000} & 0.652{\scriptsize$\pm$0.000} & 0.780{\scriptsize$\pm$0.000} & 0.640{\scriptsize$\pm$0.000} & 0.832{\scriptsize$\pm$0.000} & 0.695{\scriptsize$\pm$0.000} & 0.734{\scriptsize$\pm$0.000} & 0.721{\scriptsize$\pm$0.000} & 0.698{\scriptsize$\pm$0.000} \\
Tier-H & 0.707{\scriptsize$\pm$0.001} & 0.670{\scriptsize$\pm$0.001} & 0.802{\scriptsize$\pm$0.000} & 0.682{\scriptsize$\pm$0.001} & 0.812{\scriptsize$\pm$0.000} & 0.700{\scriptsize$\pm$0.001} & 0.765{\scriptsize$\pm$0.000} & 0.761{\scriptsize$\pm$0.000} & 0.705{\scriptsize$\pm$0.002} \\
\addlinespace[0.05cm]
\rowcolor{blue!8}
\textbf{Adaptive} & \textbf{0.703}{\scriptsize$\pm$0.000} & \textbf{0.672}{\scriptsize$\pm$0.000} & \textbf{0.810}{\scriptsize$\pm$0.000} & \textbf{0.700}{\scriptsize$\pm$0.000} & \textbf{0.835}{\scriptsize$\pm$0.000} & \textbf{0.728}{\scriptsize$\pm$0.000} & \textbf{0.766}{\scriptsize$\pm$0.000} & \textbf{0.768}{\scriptsize$\pm$0.000} & \textbf{0.710}{\scriptsize$\pm$0.001} \\
\addlinespace[0.05cm]
\midrule
\rowcolor{green!10}
$\Delta$ vs CLAP (ZS)& \textcolor{blue}{+0.175} & \textcolor{blue}{+0.130} & \textcolor{blue}{+0.270} & \textcolor{blue}{+0.126} & \textcolor{blue}{+0.148} & \textcolor{blue}{+0.172} & \textcolor{blue}{+0.158} & \textcolor{blue}{+0.202} & \textcolor{blue}{+0.158} \\
\rowcolor{green!10}
$\Delta$ vs AcuLa (ZS) & \textcolor{blue}{+0.101} & \textcolor{blue}{+0.007} & \textcolor{blue}{+0.042} & \textcolor{blue}{+0.017} & \textcolor{blue}{+0.046} & \textcolor{red}{-0.027} & \textcolor{blue}{+0.052} & \textcolor{blue}{+0.066} & \textcolor{blue}{+0.054} \\
\bottomrule
\end{tabular}
\end{adjustbox}
\end{table*}

\begin{table*}[t]
\centering
\caption{AUROC ($\uparrow$) on nine respiratory audio classification tasks. We compare \textbf{supervised} linear-probing baselines (frozen encoders with task-specific training) and \textbf{zero-shot} methods (no task-specific training). CLAP (ZS) predicts via audio--text similarity with its own pretrained encoder. \textbf{TRIAGE} uses a frozen AcuLa audio--text encoder and allocates test-time compute across three tiers: \textbf{Tier-L} (label-name cosine scoring), \textbf{Tier-M} (clinician-approved descriptor template matching with rule voting), and \textbf{Tier-H} (FAISS $k$NN report retrieval + LLM reasoning). \textbf{Adaptive} routes examples hierarchically based on confidence. $\Delta$ rows report gains of Adaptive over CLAP (ZS) and AcuLa (ZS).}
\label{tab:policy-bins-per-task}
\setlength{\tabcolsep}{7.0pt}
\renewcommand{\arraystretch}{1.15}
\small
\begin{tabular}{@{}l ccc cccc cccc@{}}
\toprule
& \multicolumn{3}{c}{\textbf{\shortstack{TL-Finalized\\(High Conf.)}}}
& \multicolumn{4}{c}{\textbf{\shortstack{TM-Finalized\\(Medium Conf.)}}}
& \multicolumn{4}{c}{\textbf{\shortstack{TH-Escalated\\(Low Conf.)}}} \\
\cmidrule(lr){2-4} \cmidrule(lr){5-8} \cmidrule(lr){9-12}
\textbf{Task} & \textbf{\%} & \textbf{TL} & \textbf{Adapt.}
& \textbf{\%} & \textbf{TL} & \textbf{Adapt.} & \textbf{Rel.$\uparrow$}
& \textbf{\%} & \textbf{TL} & \textbf{Adapt.} & \textbf{Rel.$\uparrow$} \\
\midrule
UKCOV-EX-1 & 47 & .645 & \cellcolor{white}.645 & 38 & .562 & \cellcolor{blue!35}\textbf{.690} & 23\% & 15 & .520 & \cellcolor{blue!50}\textbf{.705} & 36\% \\
UKCOV-CO-1 & 62 & .680 & \cellcolor{white}.680 & 25 & .602 & \cellcolor{blue!20}\textbf{.662} & 10\% & 13 & .566 & \cellcolor{blue!25}\textbf{.658} & 16\% \\
CVID-CO-1 & 52 & .779 & \cellcolor{white}.779 & 31 & .711 & \cellcolor{blue!30}\textbf{.804} & 13\% & 17 & .660 & \cellcolor{blue!35}\textbf{.783} & 19\% \\
CVID-CO-2 & 43 & .707 & \cellcolor{white}.707 & 37 & .640 & \cellcolor{blue!18}\textbf{.695} & 9\% & 20 & .638 & \cellcolor{blue!15}\textbf{.678} & 6\% \\
ICBHI-LS-1 & 47 & .761 & \cellcolor{white}.761 & 36 & .679 & \cellcolor{blue!35}\textbf{.797} & 17\% & 17 & .633 & \cellcolor{blue!45}\textbf{.802} & 27\% \\
COSW-CO-1 & 67 & .712 & \cellcolor{white}.712 & 21 & .690 & \cellcolor{blue!12}\textbf{.720} & 4\% & 12 & .639 & \cellcolor{blue!25}\textbf{.722} & 13\% \\
COSW-CO-2 & 36 & .745 & \cellcolor{white}.745 & 43 & .728 & \cellcolor{blue!18}\textbf{.776} & 7\% & 21 & .680 & \cellcolor{blue!22}\textbf{.750} & 10\% \\
KAUH-LS-1 & 33 & .703 & \cellcolor{white}.703 & 40 & .645 & \cellcolor{blue!25}\textbf{.728} & 13\% & 27 & .605 & \cellcolor{blue!38}\textbf{.740} & 22\% \\
RESPTR-LS-1 & 32 & .677 & \cellcolor{white}.677 & 41 & .590 & \cellcolor{blue!38}\textbf{.718} & 22\% & 27 & .545 & \cellcolor{blue!50}\textbf{.731} & 34\% \\
\midrule
\textbf{Mean} & \textbf{46} & \textbf{.712} & \cellcolor{white}\textbf{.712}
& \textbf{35} & \textbf{.646} & \cellcolor{blue!28}\textbf{.732} & \textbf{13\%}
& \textbf{19} & \textbf{.621} & \cellcolor{blue!36}\textbf{.741} & \textbf{19\%} \\
\bottomrule
\end{tabular}
\vspace{0.1cm}
\begin{flushleft}
\scriptsize
\textbf{Note:} Rel.$\uparrow$ = Relative improvement over TL baseline. Color intensity proportional to improvement magnitude. TH-Escalated shows highest mean relative gain (19\%), validating LLM reasoning value on complex cases. The TL-Finalized block omits Rel.$\uparrow$ because Adaptive and TL are identical for examples finalized at Tier-L. 
\end{flushleft}
\end{table*}

\subsection{Evaluation Protocol and Ablations}
\label{sec:eval_protocol}
We report AUROC on the held out test set for all tasks. Each experiment is repeated with five random seeds for the supervised baselines and for stochastic components of the test time pipeline, and we report mean and standard deviation.
\paragraph{Tier isolation.}
To study the effect of adaptive inference, we compare four policies that share the same backbone and data splits: Tier-L only, Tier-M only, Tier-H only, and the full adaptive router that escalates recordings based on \(c_L(x)\) and \(c_M(x)\).
\paragraph{Ablation studies.}
We conduct ablations along three axes: (i)~descriptor coverage, by masking subsets of descriptor groups in Tier-M; (ii)~retrieval depth, by varying the number of neighbors \(k \in \{1, 3, 5, 10\}\) passed to the LLM in Tier-H; and (iii)~LLM backend, by comparing Gemini 3 Pro,  gpt-oss-20, Mistral-Small-3.2-24B-Instruct, and Kimi-K2-Instruct under identical retrieval and prompt conditions. All ablations hold other components fixed to isolate each factor's contribution.

\begin{table*}[!ht]
\centering
\caption{Impact of random descriptor masking on Tier-M retrieval performance. Each task evaluated at three masking levels: 0\% (baseline), 20\%, and 50\% random descriptor removal. Performance degradation ($\Delta$) color-coded: \colorbox{green!20}{minimal} ($|\Delta| < 0.02$), \colorbox{yellow!30}{moderate} ($0.02 \leq |\Delta| < 0.05$), \colorbox{red!20}{severe} ($|\Delta| \geq 0.05$). Results based on 5 repeated maskings per rate.}
\label{tab:mask-per-task-wide-t2only}
\setlength{\tabcolsep}{6pt}
\renewcommand{\arraystretch}{1.05}
\small
\begin{tabular}{@{}llcccccc@{}}
\toprule
\textbf{Task ID} & \textbf{Classification Task} & \textbf{Baseline} & \multicolumn{2}{c}{\textbf{20\% Masking}} & \multicolumn{2}{c}{\textbf{50\% Masking}} & \textbf{Sensitivity} \\
\cmidrule(lr){3-3} \cmidrule(lr){4-5} \cmidrule(lr){6-7} \cmidrule(lr){8-8}
& & \textbf{AUROC} & \textbf{AUROC} & \textbf{$\Delta$} & \textbf{AUROC} & \textbf{$\Delta$} & \textbf{Rank} \\
\midrule
ICBHI-LS-1 & COPD (Lung sounds) & 0.832 & 0.794 & \cellcolor{yellow!30}-0.038 & 0.739 & \cellcolor{red!20}\textbf{-0.093} & High \\
UKCOV-EX-1 & COVID-19 (Exhalation) & 0.690 & 0.656 & \cellcolor{yellow!30}-0.034 & 0.607 & \cellcolor{red!20}\textbf{-0.083} & High \\
RESPTR-LS-1 & COPD Severity (Lung) & 0.698 & 0.662 & \cellcolor{yellow!30}-0.036 & 0.660 & \cellcolor{yellow!30}-0.038 & High \\
CVID-CO-1 & COVID-19 (Cough) & 0.780 & 0.763 & \cellcolor{green!20}-0.017 & 0.734 & \cellcolor{yellow!30}-0.046 & Medium \\
KAUH-LS-1 & Obstructive Disease (Lung) & 0.721 & 0.703 & \cellcolor{green!20}-0.018 & 0.690 & \cellcolor{yellow!30}-0.031 & Medium \\
COSW-CO-1 & Smoking Status (Cough) & 0.695 & 0.688 & \cellcolor{green!20}-0.007 & 0.673 & \cellcolor{yellow!30}-0.022 & Low \\
UKCOV-CO-1 & COVID-19 (Cough) & 0.652 & 0.642 & \cellcolor{green!20}-0.010 & 0.635 & \cellcolor{green!20}-0.017 & Low \\
COSW-CO-2 & Gender (Cough) & 0.734 & 0.727 & \cellcolor{green!20}-0.007 & 0.726 & \cellcolor{green!20}-0.008 & Low \\
CVID-CO-2 & Gender (Cough) & 0.640 & 0.637 & \cellcolor{green!20}-0.003 & 0.635 & \cellcolor{green!20}-0.005 & Low \\
\midrule
\multicolumn{2}{l}{\textbf{Mean Performance}} & \textbf{0.716} & \textbf{0.697} & \textbf{-0.019} & \textbf{0.678} & \textbf{-0.038} & -- \\
\multicolumn{2}{l}{\textbf{Std. Deviation}} & \textbf{0.058} & \textbf{0.053} & \textbf{0.014} & \textbf{0.047} & \textbf{0.032} & -- \\
\bottomrule
\end{tabular}
\vspace{0.1cm}
\begin{flushleft}
\scriptsize
\textbf{Note:} Sensitivity ranking based on performance drop at 50\% masking: High ($\Delta \leq -0.05$), Medium ($-0.05 < \Delta \leq -0.03$), Low ($\Delta > -0.03$). Lung sound tasks show higher sensitivity to descriptor removal than cough-based tasks, suggesting stronger reliance on semantic retrieval.
\end{flushleft}
\end{table*}

\begin{table*}[!ht]
\centering
\caption{Tier-H LLM performance versus retrieval context depth (single call, $b{=}1$). Each task evaluated with 1, 3, 5, and 8 retrieved documents. Optimal context size (highest AUROC) highlighted in bold. Gain metrics show improvement from minimal context ($d{=}1$) to optimal depth. Results demonstrate diminishing returns beyond $d{=}3$ for most tasks.}
\label{tab:docs-per-task-wide}
\setlength{\tabcolsep}{4pt}
\renewcommand{\arraystretch}{1.00}
\small
\begin{tabular}{@{}llcccccc@{}}
\toprule
\textbf{Task ID} & \textbf{Classification} & \textbf{$d{=}1$} & \textbf{$d{=}3$} & \textbf{$d{=}5$} & \textbf{$d{=}8$} & \textbf{Optimal} & \textbf{Gain} \\
 & \textbf{Task} & \textbf{(baseline)} & & & & \textbf{$d$} & \textbf{$\Delta$} \\
\midrule
\multicolumn{8}{l}{\textit{\textbf{High Context Sensitivity (Gain $>$ 0.035)}}} \\
\addlinespace[0.05cm]
RESPTR-LS-1 & COPD Severity (Lung) & 0.667 & 0.705 & 0.700 & 0.698 & \textbf{3} & \cellcolor{blue!30}+0.038 \\
CVID-CO-2 & Gender (Cough) & 0.646 & 0.682 & \textbf{0.682} & 0.680 & \textbf{3/5} & \cellcolor{blue!25}+0.036 \\
ICBHI-LS-1 & COPD (Lung sounds) & 0.785 & 0.812 & \textbf{0.818} & 0.817 & \textbf{5} & \cellcolor{blue!25}+0.033 \\
\addlinespace[0.1cm]
\multicolumn{8}{l}{\textit{\textbf{Medium Context Sensitivity (0.020 $<$ Gain $\leq$ 0.035)}}} \\
\addlinespace[0.05cm]
UKCOV-EX-1 & COVID-19 (Exhalation) & 0.680 & 0.707 & 0.705 & \textbf{0.709} & \textbf{8} & \cellcolor{blue!15}+0.029 \\
CVID-CO-1 & COVID-19 (Cough) & 0.784 & 0.802 & \textbf{0.810} & 0.805 & \textbf{5} & \cellcolor{blue!15}+0.026 \\
UKCOV-CO-1 & COVID-19 (Cough) & 0.645 & \textbf{0.670} & 0.669 & 0.666 & \textbf{3} & \cellcolor{blue!15}+0.025 \\
COSW-CO-1 & Smoking Status (Cough) & 0.681 & 0.700 & \textbf{0.705} & 0.702 & \textbf{5} & \cellcolor{blue!12}+0.024 \\
KAUH-LS-1 & Obstructive Disease (Lung) & 0.738 & \textbf{0.761} & 0.760 & 0.757 & \textbf{3} & \cellcolor{blue!12}+0.023 \\
COSW-CO-2 & Gender (Cough) & 0.746 & 0.765 & \textbf{0.769} & 0.768 & \textbf{5} & \cellcolor{blue!12}+0.023 \\
\midrule
\multicolumn{2}{l}{\textbf{Mean Performance}} & \textbf{0.708} & \textbf{0.734} & \textbf{0.735} & \textbf{0.734} & -- & \textbf{+0.028} \\
\multicolumn{2}{l}{\textbf{Optimal Context Distribution}} & 0 tasks & 4 tasks & 4 tasks & 1 task & \textbf{Mode: 3--5} & -- \\
\bottomrule
\end{tabular}
\vspace{0.1cm}
\begin{flushleft}
\scriptsize
\textbf{Note:} Mean AUROC plateaus at $d{=}3$ (0.734), with negligible improvement at $d{=}5$ (+0.001) and $d{=}8$ (-0.001). Most tasks (8/9) achieve optimal performance with 3--5 documents. Only UKCOV-EX-1 benefits from extended context ($d{=}8$). Gain magnitude correlates with task difficulty: lowest baseline performers show largest context-driven improvements.
\end{flushleft}
\end{table*}

\begin{table*}[!ht]
\centering
\caption{\textbf{Effect of Tier-L confidence cutoff $\tau_1$ on TRIAGE performance and compute.}
AUROC and tier usage (\% of samples handled by Tier-L / Tier-M / Tier-H) for $\tau_1\in\{0.30,0.45,0.60\}$.
The Tier-M$\rightarrow$Tier-H escalation threshold $\tau_2$ is fixed across all settings, and Tier-H uses the same retrieval+LLM backend throughout.}
\label{tab:gate-per-task-wide}
\setlength{\tabcolsep}{2.0pt}
\renewcommand{\arraystretch}{1.1}
\scriptsize
\begin{tabular}{@{}llcccccccccccc@{}}
\toprule
\multirow{3}{*}{\textbf{ID}} & \multirow{3}{*}{\textbf{Task}} 
& \multicolumn{4}{c}{\textbf{$\tau_1{=}0.30$}} 
& \multicolumn{4}{c}{\textbf{$\tau_1{=}0.45$}} 
& \multicolumn{4}{c}{\textbf{$\tau_1{=}0.60$}} \\
\cmidrule(lr){3-6}\cmidrule(lr){7-10}\cmidrule(lr){11-14}
& & \textbf{AUROC} & \rotatebox{90}{\textbf{\%T-L}} & \rotatebox{90}{\textbf{\%T-M}} & \rotatebox{90}{\textbf{\%T-H}}
  & \textbf{AUROC} & \rotatebox{90}{\textbf{\%T-L}} & \rotatebox{90}{\textbf{\%T-M}} & \rotatebox{90}{\textbf{\%T-H}}
  & \textbf{AUROC} & \rotatebox{90}{\textbf{\%T-L}} & \rotatebox{90}{\textbf{\%T-M}} & \rotatebox{90}{\textbf{\%T-H}} \\
\midrule
UKCOV-EX-1 & COVID (Exh.)  & 0.708 & 41 & 43 & 16 & 0.713 & 34 & 48 & 18 & 0.711 & 27 & 54 & 19 \\
UKCOV-CO-1 & COVID (Cgh.) & 0.675 & 57 & 29 & 14 & 0.678 & 47 & 37 & 16 & 0.677 & 39 & 43 & 18 \\
CVID-CO-1 & COVID (Cgh.) & 0.817 & 46 & 36 & 18 & 0.819 & 38 & 42 & 20 & 0.818 & 30 & 48 & 22 \\
CVID-CO-2 & Gender (Cgh.) & 0.702 & 36 & 43 & 21 & 0.705 & 31 & 47 & 22 & 0.703 & 23 & 54 & 23 \\
ICBHI-LS-1 & COPD (Lung) & 0.844 & 43 & 39 & 18 & 0.844 & 33 & 47 & 20 & 0.846 & 27 & 52 & 21 \\
COSW-CO-1 & Smoker (Cgh.)  & 0.727 & 63 & 24 & 13 & 0.730 & 53 & 32 & 15 & 0.728 & 43 & 40 & 17 \\
COSW-CO-2 & Gender (Cgh.)  & 0.770 & 28 & 50 & 22 & 0.772 & 24 & 53 & 23 & 0.771 & 20 & 56 & 24 \\
KAUH-LS-1 & Obstr. (Lung) & 0.775 & 25 & 46 & 29 & 0.777 & 20 & 50 & 30 & 0.779 & 16 & 53 & 31 \\
RESPTR-LS-1 & COPD Sev. & 0.717 & 24 & 47 & 29 & 0.720 & 19 & 51 & 30 & 0.721 & 17 & 52 & 31 \\
\midrule
\multicolumn{2}{l}{\textbf{Mean}} 
& 0.748 & 40.3 & 39.7 & 20.0 
& 0.751 & 33.2 & 45.2 & 21.6 
& 0.750 & 26.9 & 50.2 & 22.9 \\
\multicolumn{2}{l}{\textbf{Avg. Esc. Rate}} 
&  & \multicolumn{3}{c}{59.7\%} 
&  & \multicolumn{3}{c}{66.8\%} 
&  & \multicolumn{3}{c}{73.1\%} \\
\multicolumn{2}{l}{\textbf{Tasks w/ Best AUROC}} 
& \multicolumn{4}{c}{0 tasks} 
& \multicolumn{4}{c}{6 tasks} 
& \multicolumn{4}{c}{3 tasks} \\
\bottomrule

\end{tabular}
\end{table*}

\section{Results}

\subsection{Overall performance across tasks}

Table~\ref{tab:policy_auroc} reports AUROC on nine respiratory audio classification tasks, comparing supervised linear-probing baselines (frozen encoders with task-specific training), prior zero-shot methods, and our zero-shot tiered inference policies. All TRIAGE tiers use the frozen AcuLa audio--text encoder; CLAP (ZS) is evaluated as an external baseline with its own pretrained encoder. Among zero-shot approaches, AcuLa (ZS) is a strong baseline, while CLAP (ZS), which relies on text--audio similarity scoring, is consistently weaker across tasks. Starting from label-cosine scoring (Tier-L), adding external evidence via Tier-M and Tier-H further improves performance. Overall, the Adaptive router achieves the best zero-shot results (mean AUROC 0.744), substantially improving over CLAP (ZS) and outperforming AcuLa (ZS) on 8 of 9 tasks (the exception is COSW-CO-1).
Within our methods, accuracy increases with richer test-time information: Tier-M and Tier-H both improve over Tier-L, and Adaptive performs best on most tasks. Despite using no task-specific training data, Adaptive also compares favorably to supervised linear probing, outperforming VGGish across tasks and matching or exceeding the strongest linear-probing baseline (AcuLa) on several datasets. Additional linear-probing baselines (OPERA-CT/OPERA-CE) on the same splits are reported in Appendix~\ref{apd:linear}.

\subsection{Where adaptive routing helps: gains concentrate on uncertain cases}

Table~\ref{tab:policy-bins-per-task} stratifies test examples by the tier where Adaptive stops: \emph{TL-Finalized} (resolved at Tier-L), \emph{TM-Finalized} (resolved at Tier-M), and \emph{TH-Escalated} (escalated to retrieval-augmented LLM reasoning). For each bucket, we report its share, Tier-L AUROC on that subset, and Adaptive AUROC on the same subset.

Across tasks, \textbf{46\%} of samples are finalized at Tier-L, where Adaptive matches Tier-L (\textbf{.712} vs.\ \textbf{.712} mean AUROC), indicating routing leaves high-confidence decisions unchanged. Gains come from harder cases: for \emph{TM-Finalized} samples (\textbf{35\%}), mean AUROC rises from \textbf{.646} to \textbf{.732} (\textbf{+13\%} relative); for \emph{TH-Escalated} samples (\textbf{19\%}), it increases from \textbf{.621} to \textbf{.741} (\textbf{+19\%} relative). Thus, adaptive computation concentrates on ambiguous inputs, with the largest gains typically in \emph{TH-Escalated} (e.g., UKCOV-EX-1, RESPTR-LS-1), where retrieval-augmented LLM reasoning is most beneficial.

\subsection{Tier-M robustness: descriptor masking ablation}
Tier-M predicts labels from clinician-reviewed descriptors using sentence-bank cosine matching and a task-specific rule table. We test sensitivity to missing descriptors by randomly masking descriptor groups at inference time (0\%, 20\%, 50\%; Table~\ref{tab:mask-per-task-wide-t2only}). Mean AUROC declines from \textbf{0.716} (0\%) to \textbf{0.697} (20\%, $\Delta=-0.019$) and \textbf{0.678} (50\%, $\Delta=-0.038$). The largest 50\% drops occur for lung-sound/exhalation tasks (ICBHI-LS-1: \textbf{$-$0.093}; UKCOV-EX-1: \textbf{$-$0.083}), while cough-based tasks are comparatively stable (e.g., COSW-CO-2: \textbf{$-$0.008}; CVID-CO-2: \textbf{$-$0.005}). This supports escalating to Tier-H when descriptor evidence is missing or unreliable.

\subsection{Tier-H context scaling: retrieval depth ablation}
\label{subsec:context_depth}
We study how Tier-H performance depends on retrieval context size by varying the number of retrieved documents included in the LLM prompt ($d \in \{1, 3, 5, 8\}$) while keeping a single LLM call ($b = 1$) and all other settings fixed (Table~\ref{tab:docs-per-task-wide}).
Increasing context from $d = 1$ to a small number of documents yields consistent gains: mean AUROC rises from \textbf{0.708} at $d = 1$ to \textbf{0.734} at $d = 3$. Additional context provides limited benefit on average (\textbf{0.735} at $d = 5$ and \textbf{0.734} at $d = 8$). Most tasks achieve their best performance with \textbf{3--5} documents (8/9 tasks), with UKCOV-EX-1 as the only case that benefits from longer context ($d = 8$). Overall, these results show that Tier-H benefits from moderate retrieval context, and that prompt length beyond a few documents typically yields diminishing returns.
We provide qualitative examples of retrieved report snippets and their alignment with query audio in Appendix~\ref{apd:retrieve_example}.

\subsection{Compute--performance tradeoff: Tier-L threshold and tier distribution}
\label{subsec:tradeoff}

We examine the compute--performance tradeoff by sweeping the Tier-L cutoff $\tau_1$, which controls how often examples are finalized at Tier-L versus escalated to higher tiers (Table~\ref{tab:gate-per-task-wide}). We evaluate three settings, $\tau_1 \in \{0.30, 0.45, 0.60\}$, while keeping the Tier-M$\rightarrow$Tier-H threshold fixed.

Increasing $\tau_1$ shifts more examples to Tier-M and Tier-H. The mean fraction finalized at Tier-L drops from \textbf{40.3\%} at $\tau_1 = 0.30$ to \textbf{26.9\%} at $\tau_1 = 0.60$, and the overall escalation rate increases from \textbf{59.7\%} to \textbf{73.1\%}. Mean AUROC changes only slightly across the sweep (\textbf{0.748}, \textbf{0.751}, \textbf{0.750} for $\tau_1 = 0.30, 0.45, 0.60$, respectively). The setting $\tau_1 = 0.45$ achieves the best AUROC on the majority of tasks (6/9) with a lower escalation rate than $\tau_1 = 0.60$, illustrating that modest increases in Tier-H usage yield limited additional gains beyond an intermediate cutoff.

\section{Conclusion}
\label{sec:conclusion}



We introduced TRIAGE, a tiered zero-shot inference framework for respiratory audio classification that routes each test example through progressively richer reasoning stages based on prediction confidence, from lightweight label-cosine scoring (Tier-L), through descriptor-based matching (Tier-M), to retrieval-augmented LLM reasoning (Tier-H). Across nine diverse tasks spanning lung sounds, coughs, and exhalation recordings, our Adaptive router achieves a mean AUROC of $0.744$ without any task-specific training, outperforming prior zero-shot baselines by substantial margins and matching or exceeding supervised linear probing on several benchmarks. Crucially, the adaptive routing mechanism concentrates additional computation on ambiguous inputs: high-confidence predictions are finalized early with no loss in accuracy, while the hardest cases benefit through retrieval-augmented reasoning. These findings demonstrate that structured, confidence-aware inference can unlock strong zero-shot performance in medical audio analysis, reducing reliance on costly labeled data while providing interpretable, tiered decision pathways suitable for clinical integration.

 \section*{Acknowledgments}
This work was supported by the NWO AiNed Fellowship Grant of A.S., and in part by Google.org and the Google Cloud Research Credits program through the Gemini Academic Program. We also acknowledge the use of the Dutch National Supercomputer Snellius for essential computational tasks.  


\bibliography{main}
\clearpage

\onecolumn
\appendix

\clearpage

\section{Tier-M Descriptor Taxonomy and Class Prototypes}
\label{apd:descriptors}

In Table ~\ref{tab:t2_prototypes}, we provide a concrete example of the Tier-M descriptor system for ICBHI-LS-1 (COPD vs.\ Healthy). For each descriptor group, we list the full set of clinician-approved templates used at inference time and the prototypical descriptor choice associated with each class in our rule table (\texttt{task\_rules}). During Tier-M, a recording is first mapped to one selected template per group via cosine matching, and the resulting descriptor profile is compared against these class prototypes to produce the label decision.

\begin{table*}[!h]
\centering
\caption{Tier-2 descriptor groups and prototypical configurations for COPD vs Healthy
lung sounds in task ICBHI-LS-1. For each descriptor group we list the full option
set used in our system, and the specific option selected by the COPD
and Healthy prototypes in \texttt{task\_rules}.}
\label{tab:t2_prototypes}
\resizebox{\textwidth}{!}{%
\begin{tabular}{l p{0.52\textwidth} p{0.18\textwidth} p{0.18\textwidth}}
\toprule
\textbf{Descriptor group} &
\textbf{System option set} &
\textbf{COPD prototype} &
\textbf{Healthy prototype} \\
\midrule
Breath sound character &
normal vesicular breathing; diminished or distant breath sounds; bronchial breathing with high pitch; absent breath sounds; amphoric hollow breathing; cavernous breathing sounds; harsh breathing with increased intensity &
diminished or distant breath sounds &
normal vesicular breathing \\
\addlinespace[3pt]
Wheeze presence &
no wheeze detected; mild expiratory wheeze; moderate expiratory wheeze; severe expiratory wheeze; inspiratory wheeze present; biphasic wheeze (inspiratory and expiratory); polyphonic multiple wheeze; monophonic single wheeze &
moderate expiratory wheeze &
no wheeze detected \\
\addlinespace[3pt]
Respiratory phase timing &
normal inspiratory to expiratory ratio 1:2; prolonged expiratory phase ratio 1:3 or greater; shortened expiratory phase ratio 1:1; prolonged inspiratory phase; equal inspiratory and expiratory phases; irregular variable phase timing; rapid shallow breathing pattern &
prolonged expiratory phase ratio 1:3 or greater &
normal inspiratory to expiratory ratio 1:2 \\
\addlinespace[3pt]
Crackle characteristics &
no crackles present; fine high-pitched crackles early inspiratory; fine crackles late inspiratory; coarse low-pitched crackles early inspiratory; coarse crackles throughout inspiration; bibasilar crackles at lung bases; diffuse crackles throughout lung fields; velcro-like crackles &
coarse low-pitched crackles early inspiratory &
no crackles present \\
\addlinespace[3pt]
Respiratory effort &
normal and effortless; mildly increased effort; moderately labored; severely labored with accessory muscle use; paradoxical breathing pattern; shallow with minimal effort; gasping or air hunger pattern &
moderately labored &
normal and effortless \\
\addlinespace[3pt]
Spectral frequency profile &
normal frequency distribution 100--1000 Hz; low frequency dominance below 400 Hz; high frequency dominance above 800 Hz; broadband frequency distribution; narrow band frequency concentration; bimodal frequency peaks; irregular frequency scatter &
low frequency dominance below 400 Hz &
normal frequency distribution 100--1000 Hz \\
\bottomrule
\end{tabular}%
}
\end{table*}

\clearpage

\section{Prompt Example for Tier-H LLM Decision}
\label{apd:prompt-tierh}
At Tier-H, we query Gemini with retrieval-augmented clinical evidence to produce the final binary decision for diagnostically uncertain cases. The prompt supplies the top-$k$ retrieved report snippets as context and constrains the model to select a diagnosis from a predefined label set. To make outputs consistent and directly usable for evaluation, we enforce a strict JSON format containing only the predicted label and a brief justification, preventing extraneous explanations.

\begin{tcolorbox}[
  colframe=black,
  title=\footnotesize{Gemini Prompt for Tier-H}
]
You are a highly experienced cardiopulmonary doctor. Given the following
reports, select the most likely/probable diagnosis from the given classes
below and write very few words justification. \\

Reports:
- The presence of expiratory wheezes in the posterior lower lung fields in this
  62-year-old male with COPD suggests airway obstruction typically associated
  with this condition. \\
- In a 58-year-old female with COPD, expiratory wheezes are noted in the
  posterior right lower lung field, indicating likely airway narrowing or
  obstruction in this region. \\
- The respiratory examination of this 59-year-old male with asthma reveals
  expiratory wheezes located at the posterior right lower lung field. These
  findings may indicate airway obstruction or constriction in this region. \\

Classes: Obstructive, Healthy \\

Your output should be JSON of the following structure:
\{``result'': ..., ``justification'': ...\}. Do not provide any other explanation.
\end{tcolorbox}

\begin{tcolorbox}[
  colframe=black,
  title=\footnotesize{Example Gemini Response (JSON)}
]
\{``result'':``Obstructive'',
  ``justification":"Expiratory wheezes with COPD/asthma indicate airway obstruction.''\}
\end{tcolorbox}

\clearpage

\section{Downstream Tasks and Datasets}
\label{apd:datasets}

This section summarizes the downstream respiratory audio classification benchmarks used in our evaluation. Table~\ref{tab:dataset_overview} lists each task ID, label space, audio modality (cough, exhalation, or lung sounds), and dataset statistics (sample counts and class distributions). Tasks span three categories: respiratory disease detection, demographic classification, and COVID-19 detection.

\begin{table*}[!ht]
\centering
\caption{Downstream evaluation tasks for respiratory audio classification. All tasks use binary or multi-class classification on audio recordings from real-world clinical and crowdsourced datasets.}
\label{tab:dataset_overview}
\setlength{\tabcolsep}{4pt}
\scriptsize
\begin{tabular}{@{}llccl@{}}
\toprule
\textbf{Task ID} & \textbf{Classification Task} & \textbf{Audio Type} & \textbf{Samples} & \textbf{Class Distribution} \\
\midrule
\multicolumn{5}{l}{\textit{\textbf{COVID-19 Detection}}} \\
\addlinespace[0.05cm]
UKCOV-EX-1 & Covid vs. Non-covid & Exhalation & 2,500 & 840 / 1,660 \\
UKCOV-CO-1 & Covid vs. Non-covid & Cough & 2,500 & 840 / 1,660 \\
CVID-CO-1 & Covid vs. Non-covid & Cough & 6,175 & 547 / 5,628 \\
\midrule
\multicolumn{5}{l}{\textit{\textbf{Demographic Classification}}} \\
\addlinespace[0.05cm]
CVID-CO-2 & Female vs. Male & Cough & 7,263 & 2,468 / 4,795 \\
COSW-CO-2 & Female vs. Male & Cough & 2,496 & 759 / 1,737 \\
\midrule
\multicolumn{5}{l}{\textit{\textbf{Respiratory Disease Detection}}} \\
\addlinespace[0.05cm]
ICBHI-LS-1 & COPD vs. Healthy & Lung sounds & 828 & 793 / 35 \\
COSW-CO-1 & Smoker vs. Non-smoker & Cough & 948 & 201 / 747 \\
KAUH-LS-1 & Obstructive vs. Healthy & Lung sounds & 234 & 129 / 105 \\
RESPTR-LS-1 & COPD Severity (5-class) & Lung sounds & 504 & 72 / 60 / 84 / 84 / 204 \\
\bottomrule
\end{tabular}
\end{table*}

\section{Tier-H backend choice: LLM ablation}
\label{apd:llm_ablation}

We replace the LLM used in Tier-H while keeping the encoder, retrieval database, and prompt fixed ($d{=}3$, same budget $b$; Table~\ref{tab:llm-swap-per-task-wide}). Gemini~3~Pro achieves the best AUROC on all nine tasks, with a mean AUROC of \textbf{0.734}. Kimi-K2 is second (\textbf{0.711}), followed by gpt-oss (\textbf{0.695}) and Mistral-Small (\textbf{0.689}). Backend choice therefore affects absolute Tier-H performance under matched inference cost, and we use Gemini~3~Pro as the default Tier-H backend in the remaining experiments.

\begin{table*}[!ht]
\caption{\textbf{Tier-H LLM backend ablation.} Per-task AUROC when swapping the Tier-H LLM while keeping retrieval and prompting fixed ($d{=}3$, same budget $b$). Backends: Gemini~3~Pro \citep{gemini3pro_modelcard}, gpt-oss-20b \citep{openai2025gptoss120bgptoss20bmodel}, Mistral-Small-3.2-24B-Instruct \citep{mistralai_mistral_small_3_2_24b_instruct_2506}, and Kimi-K2-Instruct \citep{moonshotai_kimi_k2_instruct}.}
\centering
\begin{adjustbox}{max width=\textwidth,center}
\footnotesize
\begin{tabular}{ll|cccc}
\toprule
\multirow{2}{*}{\textbf{ID}} & \multirow{2}{*}{\textbf{Task}} 
& \multicolumn{4}{c}{\textbf{AUROC ($\uparrow$)}} \\
\cmidrule(lr){3-6}
& & \textbf{Gemini 3 Pro} & \textbf{gpt-oss} & \textbf{Mistral-Small} & \textbf{Kimi-K2} \\
\midrule
UKCOV-EX-1  & Covid (Exhale)        & 0.707 & 0.686 & 0.660 & 0.683 \\
UKCOV-CO-1  & Covid (Cough)         & 0.670 & 0.643 & 0.638 & 0.659 \\
CVID-CO-1   & Covid (Cough)         & 0.802 & 0.756 & 0.752 & 0.767 \\
CVID-CO-2   & Gender (Cough)        & 0.682 & 0.654 & 0.648 & 0.668 \\
ICBHI-LS-1  & COPD (Lung)           & 0.812 & 0.755 & 0.757 & 0.776 \\
COSW-CO-1   & Smoker (Cough)        & 0.700 & 0.664 & 0.663 & 0.685 \\
COSW-CO-2   & Gender (Cough)        & 0.765 & 0.716 & 0.718 & 0.742 \\
KAUH-LS-1   & Obstructive (Lung)    & 0.761 & 0.713 & 0.710 & 0.739 \\
RESPTR-LS-1 & COPD severity (Lung)  & 0.705 & 0.666 & 0.659 & 0.676 \\
\bottomrule
\end{tabular}
\end{adjustbox}
\label{tab:llm-swap-per-task-wide}
\end{table*}

\section{Additional linear-probing baselines}
\label{apd:linear}
Table~\ref{tab:oct_oce_appendix} reports the per-task linear-probing AUROC of OPERA-CT and OPERA-CE \citep{zhang2024towards} on the same train/test splits as Table~\ref{tab:policy_auroc}. For completeness, we also include the corresponding tiered inference results (TL--TH and Adaptive) on the same tasks.

\begin{table}[!ht]
\centering
\caption{Appendix: AUROC ($\uparrow$) for the omitted linear-probing baselines OPERA-CT and OPERA-CE, alongside our three-tier inference policies (same splits as Table~\ref{tab:policy_auroc}).}
\label{tab:oct_oce_appendix}
\setlength{\tabcolsep}{4.0pt}
\renewcommand{\arraystretch}{1.10}
\scriptsize
\begin{tabular}{l | c c | c c c c}
\toprule
\textbf{ID} & \textbf{OCT} & \textbf{OCE} & \textbf{TL-only} & \textbf{TM-only} & \textbf{TH-only} & \textbf{Adaptive} \\
\midrule
UKCOV-EX-1   & 0.586  & 0.551  & 0.593  & 0.690 & 0.707 & 0.703 \\
UKCOV-CO-1   & 0.701 & 0.629 & 0.627  & 0.652& 0.670 & 0.672  \\
CVID-CO-1    & 0.578 & 0.566 & 0.722  & 0.780 & 0.802  & 0.810\\
CVID-CO-2    & 0.795  & 0.721  & 0.668  & 0.640 & 0.682  & 0.700 \\
ICBHI-LS-1   & 0.855  & 0.872 & 0.706 & 0.832 & 0.812 & 0.835  \\
COSW-CO-1    & 0.685  & 0.674 & 0.717  & 0.695  & 0.700 & 0.728  \\
COSW-CO-2    & 0.874  & 0.801 & 0.716 & 0.734  & 0.765  & 0.766  \\
KAUH-LS-1    & 0.722 & 0.741  & 0.670  & 0.721 & 0.761& 0.768  \\
RESPTR-LS-1  & 0.625 & 0.683 & 0.610 & 0.698 & 0.705  & 0.710  \\
\bottomrule
\end{tabular}
\end{table}

\clearpage

\section{Qualitative Retrieval Examples}
\label{apd:retrieve_example}

This section presents qualitative retrieval results from the Tier-H stage. For each query auscultation clip, FAISS retrieves the top-3 nearest clinical reports in the shared audio--text embedding space. The retrieved reports serve as supporting context for the Tier-H decision module, and they also provide an interpretable view of what the retrieval stage considers most similar to the query.

\begin{figure*}[!ht]
  \captionsetup{position=top,font=small}
  \centering
  \caption{Top-3 clinical reports retrieved for auscultation clips (Tier-H).
  Left: query spectrogram and its reference (or generated) report.
  Right: the three closest reports returned by FAISS in the shared embedding space.}
  \includegraphics[width=\textwidth]{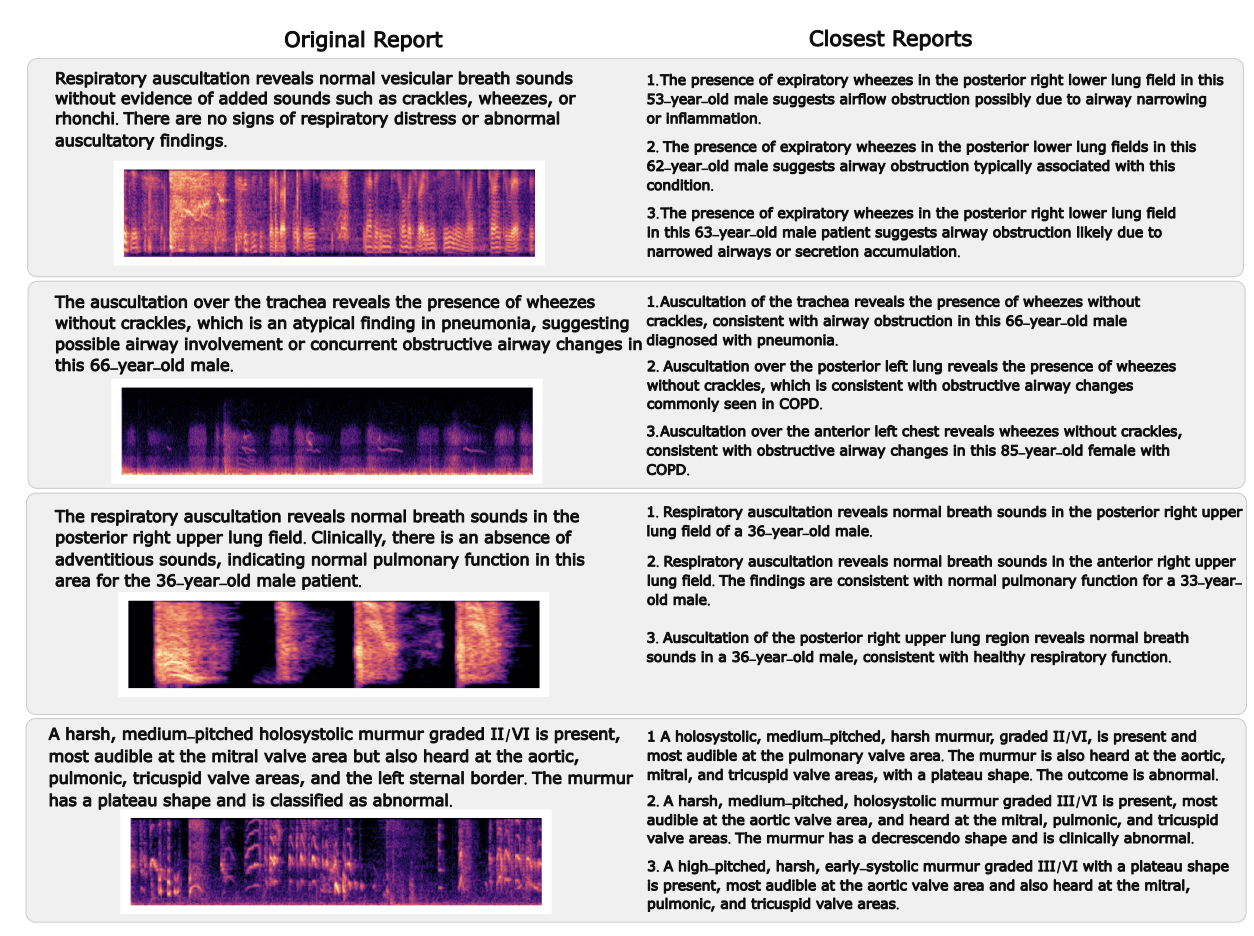}
  \label{fig:closest_report}
\end{figure*}

\end{document}